\input harvmac
\input epsf

\newcount\figno
\figno=0

\def\fig#1#2#3{
\par\begingroup\parindent=0pt\leftskip=1cm\rightskip=1cm\parindent=0pt
\baselineskip=11pt \global\advance\figno by 1 \midinsert
\epsfxsize=#3 \centerline{\epsfbox{#2}} \vskip 12pt {\bf Figure\
\the\figno: } #1\par
\endinsert\endgroup\par
}
\def\figlabel#1{\xdef#1{\the\figno}}

\newcount\yearltd\yearltd=\year\advance\yearltd by 0
\def\IR{\relax{\rm I\kern-.18em R}}
\noblackbox
\def\co{{\cal O}}


\lref\MaldacenaRE{
  J.~M.~Maldacena,
  ``The large N limit of superconformal field theories and supergravity,''
  Adv.\ Theor.\ Math.\ Phys.\  {\bf 2}, 231 (1998)
  [Int.\ J.\ Theor.\ Phys.\  {\bf 38}, 1113 (1999)]
  [arXiv:hep-th/9711200].
}

\lref\WittenQJ{
  E.~Witten,
  ``Anti-de Sitter space and holography,''
  Adv.\ Theor.\ Math.\ Phys.\  {\bf 2}, 253 (1998)
  [arXiv:hep-th/9802150].
}

\lref\GubserBC{
  S.~S.~Gubser, I.~R.~Klebanov and A.~M.~Polyakov,
  ``Gauge theory correlators from non-critical string theory,''
  Phys.\ Lett.\ B {\bf 428}, 105 (1998)
  [arXiv:hep-th/9802109].
}

\lref\KlebanovHH{
  I.~R.~Klebanov and E.~Witten,
  ``Superconformal field theory on threebranes at a Calabi-Yau  singularity,''
  Nucl.\ Phys.\ B {\bf 536}, 199 (1998)
  [arXiv:hep-th/9807080].
}

\lref\WittenUA{
  E.~Witten,
  ``Multi-trace operators, boundary conditions, and AdS/CFT correspondence,''
  arXiv:hep-th/0112258.
}

\lref\BerkoozUG{
  M.~Berkooz, A.~Sever and A.~Shomer,
  ``Double-trace deformations, boundary conditions and spacetime
  singularities,''
  JHEP {\bf 0205}, 034 (2002)
  [arXiv:hep-th/0112264].
}

\lref\AharonySH{
  O.~Aharony, M.~Berkooz and B.~Katz,
  ``Non-local effects of multi-trace deformations in the AdS/CFT
  correspondence,''
  JHEP {\bf 0510}, 097 (2005)
  [arXiv:hep-th/0504177].
}

\lref\MaldacenaKR{
  J.~M.~Maldacena,
  ``Eternal black holes in Anti-de-Sitter,''
  JHEP {\bf 0304}, 021 (2003)
  [arXiv:hep-th/0106112].
}

\lref\AharonyPA{
  O.~Aharony, M.~Berkooz and E.~Silverstein,
  ``Multiple-trace operators and non-local string theories,''
  JHEP {\bf 0108}, 006 (2001)
  [arXiv:hep-th/0105309].
}

\lref\PorratiCP{
  M.~Porrati,
  ``No van Dam-Veltman-Zakharov discontinuity in AdS space,''
  Phys.\ Lett.\ B {\bf 498}, 92 (2001)
  [arXiv:hep-th/0011152].
}
\lref\KoganUY{
  I.~I.~Kogan, S.~Mouslopoulos and A.~Papazoglou,
  ``The m $\to$ 0 limit for massive graviton in dS(4) and AdS(4): How to
  circumvent the van Dam-Veltman-Zakharov discontinuity,''
  Phys.\ Lett.\ B {\bf 503}, 173 (2001)
  [arXiv:hep-th/0011138].
}

\lref\PorratiSA{
  M.~Porrati,
  ``Higgs phenomenon for the graviton in AdS space,''
  Mod.\ Phys.\ Lett.\ A {\bf 18}, 1793 (2003)
  [arXiv:hep-th/0306253].
}

\lref\DimopoulosUI{
  S.~Dimopoulos, S.~Kachru, N.~Kaloper, A.~E.~Lawrence and E.~Silverstein,
  ``Small numbers from tunneling between brane throats,''
  Phys.\ Rev.\ D {\bf 64}, 121702 (2001)
  [arXiv:hep-th/0104239].
}

\lref\DimopoulosQD{
  S.~Dimopoulos, S.~Kachru, N.~Kaloper, A.~E.~Lawrence and E.~Silverstein,
  ``Generating small numbers by tunneling in multi-throat  compactifications,''
  Int.\ J.\ Mod.\ Phys.\ A {\bf 19}, 2657 (2004)
  [arXiv:hep-th/0106128].
}

\lref\RandallEE{
  L.~Randall and R.~Sundrum,
  ``A large mass hierarchy from a small extra dimension,''
  Phys.\ Rev.\ Lett.\  {\bf 83}, 3370 (1999)
  [arXiv:hep-ph/9905221].
}

\lref\RandallVF{
  L.~Randall and R.~Sundrum,
  ``An alternative to compactification,''
  Phys.\ Rev.\ Lett.\  {\bf 83}, 4690 (1999)
  [arXiv:hep-th/9906064].
}

\lref\PorratiGX{
  M.~Porrati,
  ``Mass and gauge invariance. IV: Holography for the Karch-Randall model,''
  Phys.\ Rev.\ D {\bf 65}, 044015 (2002)
  [arXiv:hep-th/0109017].
}

\lref\PorratiDB{
  M.~Porrati,
  ``Higgs phenomenon for 4-D gravity in anti de Sitter space,''
  JHEP {\bf 0204}, 058 (2002)
  [arXiv:hep-th/0112166].
}

\lref\DuffWH{
  M.~J.~Duff, J.~T.~Liu and H.~Sati,
  ``Complementarity of the Maldacena and Karch-Randall pictures,''
  Phys.\ Rev.\ D {\bf 69}, 085012 (2004)
  [arXiv:hep-th/0207003].
}

\lref\KarchCT{
  A.~Karch and L.~Randall,
  ``Locally localized gravity,''
  JHEP {\bf 0105}, 008 (2001)
  [arXiv:hep-th/0011156].
}

\lref\AharonyQF{
  O.~Aharony, O.~DeWolfe, D.~Z.~Freedman and A.~Karch,
  ``Defect conformal field theory and locally localized gravity,''
  JHEP {\bf 0307}, 030 (2003)
  [arXiv:hep-th/0303249].
}

\lref\MaldacenaSS{
  J.~M.~Maldacena, G.~W.~Moore and N.~Seiberg,
  ``D-brane charges in five-brane backgrounds,''
  JHEP {\bf 0110}, 005 (2001)
  [arXiv:hep-th/0108152].
}

\lref\KimEZ{
  H.~J.~Kim, L.~J.~Romans and P.~van Nieuwenhuizen,
  ``The Mass Spectrum Of Chiral N=2 D = 10 Supergravity On S**5,''
  Phys.\ Rev.\ D {\bf 32}, 389 (1985).
}

\lref\OsbornTT{
  H.~Osborn and A.~C.~Petkou,
  ``Implications of conformal invariance in field theories for general
  dimensions,''
  Annals Phys.\  {\bf 231}, 311 (1994)
  [arXiv:hep-th/9307010].
}

\lref\AharonyBX{
  O.~Aharony, A.~Hanany, K.~A.~Intriligator, N.~Seiberg and M.~J.~Strassler,
  ``Aspects of N = 2 supersymmetric gauge theories in three dimensions,''
  Nucl.\ Phys.\ B {\bf 499}, 67 (1997)
  [arXiv:hep-th/9703110].
}

\lref\danbitensors{  E.~D'Hoker, D.~Z.~Freedman, S.~D.~Mathur,
A.~Matusis and L.~Rastelli,
  ``Graviton and gauge boson propagators in AdS(d+1),''
  Nucl.\ Phys.\ B {\bf 562}, 330 (1999)
  [arXiv:hep-th/9902042].
}

\lref\AllenWD{
  B.~Allen and T.~Jacobson,
  ``Vector Two Point Functions In Maximally Symmetric Spaces,''
  Commun.\ Math.\ Phys.\  {\bf 103}, 669 (1986).
}

\lref\allenturyn{
  B.~Allen and M.~Turyn,
  ``An Evaluation Of The Graviton Propagator In De Sitter Space,''
  Nucl.\ Phys.\ B {\bf 292}, 813 (1987).
}

\lref\FreedmanTZ{
  D.~Z.~Freedman, S.~D.~Mathur, A.~Matusis and L.~Rastelli,
  ``Correlation functions in the CFT($d$)/AdS($d+1$) correspondence,''
  Nucl.\ Phys.\ B {\bf 546}, 96 (1999)
  [arXiv:hep-th/9804058].
}

\lref\KlebanovJA{
  I.~R.~Klebanov and A.~M.~Polyakov,
  ``AdS dual of the critical O(N) vector model,''
  Phys.\ Lett.\ B {\bf 550}, 213 (2002)
  [arXiv:hep-th/0210114].
}

\lref\LiuBU{
  H.~Liu and A.~A.~Tseytlin,
   ``D = 4 super Yang-Mills, D = 5 gauged supergravity, and D = 4 conformal
  supergravity,''
  Nucl.\ Phys.\ B {\bf 533}, 88 (1998)
  [arXiv:hep-th/9804083].
}

\lref\SeibergPQ{
  N.~Seiberg,
  ``Electric - magnetic duality in supersymmetric nonAbelian gauge theories,''
  Nucl.\ Phys.\ B {\bf 435}, 129 (1995)
  [arXiv:hep-th/9411149].
}

\lref\WeinbergKQ{
  S.~Weinberg and E.~Witten,
  ``Limits On Massless Particles,''
  Phys.\ Lett.\ B {\bf 96}, 59 (1980).
}

\lref\ofertalk{ {\tt
http://www.damtp.cam.ac.uk/estg06/talks/aharony/index.html}.}

\lref\ElitzurKZ{
  S.~Elitzur, A.~Giveon, M.~Porrati and E.~Rabinovici,
   ``Multitrace deformations of vector and adjoint theories and their
  holographic duals,''
  JHEP {\bf 0602}, 006 (2006)
  [arXiv:hep-th/0511061].
}

\lref\GabadadzeJM{
  G.~Gabadadze, L.~Grisa and Y.~Shang,
  ``Resonance in asymmetric warped geometry,''
  arXiv:hep-th/0604218.
}

\lref\KiritsisHY{
  E.~Kiritsis,
   ``Product CFTs, gravitational cloning, massive gravitons and the space of
  gravitational duals,''
  arXiv:hep-th/0608088.
}


\Title{\vbox{\baselineskip12pt\hbox{hep-th/0608089,
WIS/09/06-AUG-DPP}}} {\vbox{{\centerline{The CFT/AdS correspondence,
massive gravitons} \vskip4pt \centerline{and a connectivity index
conjecture}}}}

\centerline{Ofer Aharony$^{1}$, Adam B. Clark$^{2}$, and Andreas
Karch$^2$}
\bigskip

\centerline{\sl $^1$ Department of Particle Physics} \centerline{\sl
The Weizmann Institute of Science, Rehovot 76100, Israel}

\centerline{\sl $^2$ Department of Physics, University of
Washington, Seattle, WA 98195, USA}

\vskip .3in
We discuss the general question of which conformal field theories
have dual descriptions in terms of quantum gravity theories on
anti-de Sitter space. We analyze in detail the case of a deformed
product of $n$ conformal field theories (each of which has a gravity
dual), and we claim that the dual description of this is by a
quantum gravity theory on a union of $n$ anti-de Sitter spaces,
connected at their boundary (by correlations between their boundary
conditions). On this union of spaces, $(n-1)$ linear combinations of
gravitons obtain a mass, and we compute this mass both from the
field theory and from the gravity sides of the correspondence,
finding the same result in both computations. This is the first
example in which a graviton mass in the bulk of anti-de Sitter space
arises continuously by varying parameters. The analysis of these
deformed product theories leads us to suggest that field theories
may be generally classified by a ``connectivity index'',
corresponding to the number of components (connected at the
boundary) in the space-time of the dual gravitational background. In
the field theory this index roughly counts the number of independent
gauge groups, but we do not have a precise general formula for the
index.

 \vskip .1in

\smallskip
\Date{August 2006}


\vfill \eject

\newsec{Introduction}

The AdS/CFT correspondence \MaldacenaRE\ implies that any theory of
quantum gravity (such as string theory) on anti-de Sitter (AdS)
space is dual to a conformal field theory (CFT). More generally, any
theory of quantum gravity on a space which has an asymptotic
boundary where it approaches a (possibly warped) product of anti-de
Sitter space with another space is equivalent to a field theory
which is conformal at high energies. This can be seen simply by
computing the correlation functions of local operators in such a
theory using the methods of \refs{\GubserBC,\WittenQJ} and noticing
that they obey the usual requirements for correlation functions in a
field theory. Of course, in general it is not known how to write
down a Lagrangian formulation of the dual field theory (or of a
theory that flows to it), and it is not even clear that such a
formulation should exist. But still, the dual field theory can be
implicitly (and presumably uniquely) defined through its correlation
functions, which we can compute if we understand the corresponding
theory of quantum gravity.

It is natural to ask whether this duality can be used also in the
opposite direction -- namely, whether any field theory is dual to a
theory of quantum gravity (on some asymptotically AdS space, since
standard local field theories are conformal at high energies)\foot{ Of
course, in most cases this quantum gravity theory would be highly
curved, with no semi-classical approximation, for instance since the
field theory would not have any separation between the dimensions of
operators with spin 2 or less and the dimensions of higher-spin
operators. Since in general we do not have any independent definition
of quantum gravity on highly curved spaces, another way to phrase the
question is whether any field theory can be used as a definition of a
theory of quantum gravity on asymptotically AdS space.}. At first
sight one might think that the answer must be positive, since
otherwise there would be a strange division of the space of field
theories into two sets -- the ones which have a quantum gravity dual
and the ones which do not. However, we would like to argue that the
answer is negative, and that the space of quantum field theories is
indeed divided into classes, such that only one class of quantum field
theories is dual to quantum gravity on (asymptotically) AdS space.

Our argument will be based on considering products of $n$ conformal
field theories, each of which is dual to some quantum gravity theory
on AdS space, and deforming them by products of operators from the
different CFTs in a way which couples them together\foot{The
generalization to deformations of non-conformal field theories is
straightforward.}. We will claim that the dual of such a deformed
product theory is not given by a quantum gravity theory on AdS
space, but rather on a union of $n$ AdS spaces, whose
boundaries are all identified together (in the sense that the
boundary conditions on the different AdS spaces are correlated to
each other). We will construct this picture in the limit where we
can describe the quantum gravity theory semi-classically (as a
theory of weakly coupled fields living on the AdS spaces), and
verify in detail that it gives a consistent description of the
deformed product field theory. We conjecture that the same picture
is true more generally, even beyond the semi-classical gravity
approximation.

It seems reasonable to assume that if some field theory has a dual
description in terms of a quantum gravity theory on a sum of $n$ AdS
spaces, it cannot also have a dual description as a theory living on
a single AdS space; however, it is of course difficult to be sure of
such a statement, since we do not understand quantum gravity beyond
the semi-classical limit, and it is possible that the same theory
could be described either as living on $n$ AdS spaces or as living
on a single AdS space, despite the fact that the topologies of these
two descriptions are different near the boundary. If our assumption
is indeed true, it suggests that quantum field theories may be
characterized by a ``connectivity index'' $n$, which counts the
number of separate components in their quantum gravity dual
description; schematically this index corresponds to the number of
independent gauge groups in the theory (by independent gauge groups
we mean groups such that no field is charged under more than one
group, and such that every group has at least one field charged
under it, so that, for example, an $SU(N)\times SU(N)$ theory with a
bi-fundamental field counts as having one independent gauge group;
the gauge groups can be continuous groups or discrete groups as in
sigma models on orbifolds). We claim that there exist theories with
$n=0$ (no non-trivial gauge symmetry) which do not have a dual
gravitational description, since the energy-momentum tensor in these
theories is not an independent operator, so there is no fundamental
dual bulk graviton that can be associated with a diffeomorphism
symmetry. Theories with $n=1$ include all the known theories with a
gravity dual. Theories with higher values of $n$ can be constructed
by (deformations of) direct products of theories with $n=1$, and we
claim that they correspond to a theory of quantum gravity on a sum
of $n$ asymptotically-AdS spaces, which are connected at their
boundary. We do not know how to define $n$ directly in the field
theory; this may be due to our lack of imagination, or it may mean
that our assumption is wrong and $n$ is not really a well-defined
index beyond the semi-classical limit.


We begin in section 2 by examining in detail a product of two
conformal field theories and its deformations, and how they are
described in the dual gravitational picture. The generalization to a
product of more field theories and to theories with no conformal
invariance is straightforward. In section 3 we briefly discuss the
case with $n=0$, and in section 4 we discuss further the
``connectivity index'' and its properties. We end in section 5 with
a summary of our results and conclusions.

\newsec{The dual of a deformed product of field theories}

\subsec{General description}

Let us consider two conformal field theories in $d$ dimensions, each
of which is dual to a string theory on a product of $AdS_{d+1}$
times some space ${\cal M}$. We consider the case where this dual
string theory has a good semi-classical gravity limit (namely all
radii of curvature are very large). The dual description of the
product of the two theories is obviously given by the product of the
string theories on the two spaces, or equivalently by string theory
on the disjoint union of the two spaces (which do not talk to each
other).

On the field theory side, we can deform such a product by adding to
its Lagrangian a term
\eqn\deformation{h \int d^dx {\cal O}_1(x) {\cal O}_2(x),}
where ${\cal O}_1$ is an operator in the first CFT and ${\cal O}_2$
an operator in the second. Such a deformation could be relevant or
marginal, and in some cases it can even be exactly marginal (for
instance, we can consider $J_1 {\bar J_2}$ deformations when $d=2$
and $J$ ($\bar J$) is a holomorphic (anti-holomorphic) global $U(1)$
current, or we can consider a product of Klebanov-Witten $d=4$ CFTs
\KlebanovHH\ and deform them by $\int d^2\theta h^{ijkl}
\tr(A_i^{(1)} B_j^{(1)}) \tr(A_k^{(2)} B_l^{(2)})$, which is exactly
marginal for an appropriate choice of the coefficients $h^{ijkl}$).
For simplicity let us consider the exactly marginal case, where we
have a CFT for every value of $h$, which we will denote by $CFT_h$;
our results may be easily generalized also to non-marginal
deformations.

Naively, one might think that for non-zero values of $h$ this
$CFT_h$ (which is no longer a direct product) should be dual to some
theory of quantum gravity on a single AdS space. One argument for
this is that the original product theory had two separate conserved
energy-momentum tensors $T^{(1)}_{mn}$ and $T^{(2)}_{mn}$, while the
deformed product only has one conserved energy-momentum tensor
(which in simple cases is given by, to leading order in $h$,
$T^{(1)}_{mn}+T^{(2)}_{mn}-h\, \eta_{mn} {\cal O}_1 {\cal O}_2$).
The dual of the product theory had two massless gravitons
corresponding to $T^{(1)}$ and $T^{(2)}$, while in the deformed
theory one would expect to remain with one massless graviton while
the other graviton would obtain a mass (related to the anomalous
dimension of the non-conserved combination of $T^{(1)}$ and
$T^{(2)}$; we will discuss this in more detail below). So, one might
think that the dual theory lives on a single space. However, if we
try to write the theory on a single space we immediately run into
the problem that the two spaces ${\cal M}_1$ and ${\cal M}_2$ may be
different, so there is no natural ten-dimensional space on which to
define $CFT_h$. This problem becomes even more serious if we try to
couple two non-conformal field theories which are dual to
asymptotically AdS spaces, since the two asymptotically AdS spaces
corresponding to the two field theories may be completely different
in the bulk, and there is no natural way to identify them.

We would like to suggest a different interpretation. In the standard
AdS/CFT correspondence, deformations of the Lagrangian are described
by changing boundary conditions for the fields. ``Single-trace''
deformations (deformations by an operator which is dual to a single
field in the bulk) involve changing the boundary condition on the
non-normalizable mode of the corresponding field, while ``multi-trace''
deformations (similar to \deformation) involve \refs{\WittenUA,
\BerkoozUG} changes in the boundary
conditions which mix the coefficients of the non-normalizable and the
normalizable modes of the fields involved; for example, the deformation
\deformation\ in a single CFT would imply that the coefficient near the
boundary of the non-normalizable mode of the field $\phi_1$ dual to
${\cal O}_1$ must be equal to $h*(2\Delta_2-d)$ times the
coefficient of the normalizable mode of $\phi_2$ (dual to ${\cal
O}_2$ of dimension $\Delta_2$), and vice versa\foot{This picture
makes sense when we have a semi-classical limit of the gravity
theory in which the bulk fields are weakly coupled and we know what
we mean by single-particle and multi-particle states and by boundary
conditions on fields.}. Since the deformation is described purely in
terms of boundary conditions, if we think of it in the two-CFT case,
we do not necessarily need the two fields (corresponding to
operators in the two CFTs) to live on the same space -- it is enough
if they share the same boundary. Note that the boundary of the
gravity dual of any theory which is conformal in the UV is always
the same, and looks like the $\IR^{d}$ (or $S^{d-1}\times \IR$)
boundary of $AdS_{d+1}$ (the compact space ${\cal M}$ shrinks to
zero near the boundary in the natural field theory units), so it is
always possible to identify the boundaries of the spaces dual to the
two CFTs. We suggest that the proper way to think about the gravity
dual of the product of two CFTs is as living on the sum of the two
space-times (one for each CFT) but with the boundary
identified\foot{Note that this is completely different from the case
of a single space-time with more than one boundary, which is
relevant (for instance) for the eternal AdS black hole
\MaldacenaKR.}; when the product is undeformed the identification of
the boundaries has no effect, but when we deform we can then
implement the deformation
\deformation\ by an appropriate change in the boundary conditions of the
fields living on the two space-times. Note that this procedure
correctly implements the breaking of the symmetry from a product of
two conformal groups when $h=0$ (and the identification has no
effect) to a single conformal group when $h$ is non-zero. Note also
that, as for any multi-trace deformations, the changed boundary
conditions are generally non-local on the compact spaces ${\cal M}$
\refs{\AharonyPA,\AharonySH} (and they also seem to be non-local on
the worldsheet of the corresponding string theory \AharonyPA), but
this is not inconsistent with the locality and causality of $CFT_h$.

While this picture of two spaces connected at their boundary may
seem rather arbitrary, it actually arises naturally in the AdS/CFT
correspondence when we have a flow from a single CFT to a product of
two CFTs. The prototypical example is the $d=4$ ${\cal N}=4$ $SU(N)$
SYM theory, at a point on its Coulomb branch at which the gauge
group is broken to $SU(N_1)\times SU(N-N_1)\times U(1)$. The gravity
dual of this theory is exactly known; it is asymptotically a single
AdS space, but in the interior of the space there are two
``throats'', one for each non-Abelian gauge group factor, and the
low-energy dynamics (below the mass scale of the W-bosons charged
under both $SU(K)$ factors) is a product of two CFTs, one in each
``throat''. Here the two low-energy theories are decoupled\foot{Up
to various irrelevant couplings, as discussed for instance in
\refs{\DimopoulosUI,\DimopoulosQD}.}, and it is not clear that the
two ``throats'' share a boundary, but it seems likely that if we
would deform the theory by a deformation that in the low-energy CFT
would take the form
\deformation, it would look like a shared-boundary interaction of
the form described above.

In the special case where we are deforming a product of CFTs by an
exactly marginal deformation, so that symmetry considerations imply
that the two spaces remain AdS spaces also after the deformation, we
could identify the two AdS spaces if we want, and obtain two
decoupled theories on a single AdS space which only talk to each
other through the boundary conditions related to
\deformation\ (if the compact spaces ${\cal M}_1$ and ${\cal M}_2$
are the same we could even identify the full ten dimensional spaces
if we want). This ``folded'' picture is, of course, completely
equivalent to the ``unfolded'' picture described above involving two
spaces connected at the boundary, except that in the ``folded''
picture we do not explicitly exhibit the two diffeomorphism
symmetries associated with the two spaces, but rather we work in a
gauge where the spaces are identified. The ``folded'' picture may be
useful for performing computations (which can sometimes be reduced
to computations which have already been done for theories living on
a single AdS space), but we stress that it is only available in very
special cases for which the spaces on the two sides are the same,
and there does not seem to be an analogous picture for a coupling
between two non-conformal theories which live on different spaces.
Thus, we view the ``unfolded'' picture involving a union of two
spaces as the more basic one.

The terminology which we were using for the gravitational
description of the deformation above depended on having a
semi-classical limit of the gravity theory, in which there were
well-defined fields to which one could assign boundary conditions.
In the absence of such a semi-classical limit it is not clear how to
phrase deformations in terms of boundary conditions; for example, in
a generic field theory there is no distinction between
``single-trace'' and ``multi-trace'' operators (and, indeed, even
when there is a semi-classical large $N$ limit, the two types of
operators mix together at finite $N$).
However, just as we believe that the AdS/CFT correspondence makes
sense even for finite $N$ (say, for the $SU(2)$ ${\cal N}=4$
super-Yang-Mills theory, which is very far from the semi-classical
large $N$ limit), we conjecture that the general claims above are
true even when there is no semi-classical limit. So, we conjecture
in general that a field theory which is a deformed product of $n$
independent CFTs is dual to a theory of quantum gravity on a space
with $n$ asymptotically-AdS regions connected at their boundary (the
generalization of the discussion above to products of $n$ CFTs
rather than two is straightforward).

We begin in subsection 2.2 by reviewing some facts about massive
gravity on AdS space, and how it can arise from a deformed product
of CFTs as described above. The rest of the section is devoted to a
detailed computation of the mass of the massive graviton resulting
from the deformation \deformation, both on the field theory side and
on the gravity side, in order to test our gravitational description
of this deformation.

\subsec{Massive gravity in anti-de Sitter space}

It is well-known that on anti-de Sitter space, unlike in flat space,
there is no discontinuity separating massive spin 2 fields from
massless spin 2 fields \refs{\PorratiCP,\KoganUY}.
Thus, it is possible for the graviton to
continuously acquire a mass, in a way which is completely analogous to
the Higgs mechanism, see \PorratiSA\ for a review.

In the usual Higgs mechanism in anti-de Sitter space, a massless
vector field and a massless scalar field join together to form a
massive vector field. This has a simple interpretation in the dual
field theory. Before Higgsing, the massless vector field maps to a
conserved current $J_{\mu}$ of dimension $(d-1)$, $\del^{\mu}
J_{\mu} = 0$, while the massless scalar field maps to an operator
${\cal O}$ of dimension $d$. The Higgs mechanism in the bulk is
reflected on the boundary by the current no longer being conserved,
$\del^{\mu} J_{\mu} \propto {\cal O}$, explicitly showing that the
vector operator and the scalar operator join together into a single
multiplet of the conformal algebra.

Similarly, a massless graviton in the bulk is dual to a conserved
spin 2 operator of dimension $d$, $\del^{\mu} T_{\mu \nu} = 0$. When
the diffeomorphism symmetry generated by this spin 2 operator is
broken, we have a relation of the form $\del^{\mu} T_{\mu \nu}
\propto K_{\nu}$ for some vector operator $K_{\nu}$ which joins
together with the spin 2 operator $T_{\mu \nu}$ in a single
multiplet of the conformal algebra. In the bulk this means that the
vector field dual to $K_{\nu}$ joins together with the graviton dual
to $T_{\mu \nu}$ to form a single massive spin 2 field (of course,
as in the previous case, once $T_{\mu \nu}$ is not conserved, the
conformal algebra implies that it must acquire an anomalous
dimension which is mapped to the bulk mass). An interesting
difference from the previous case is that the vector field which is
swallowed is not massless. In the previous paragraph the operator
${\cal O}$ had to have dimension $d$ before the Higgsing for a
relation of the form $\del^{\mu} J_{\mu} \propto {\cal O}$ to make
sense; similarly, the relation $\del^{\mu} T_{\mu \nu} \propto
K_{\nu}$ implies that $K_{\nu}$ must have (before the deformation)
dimension $(d+1)$, corresponding to a vector field of mass $m^2
R_{AdS}^2 = 2 d$ (recall that a massless vector field in AdS maps to
a vector operator of dimension $(d-1)$). Nevertheless, the conformal
algebra joins together this massive vector representation with the
massless spin 2 representation to form the massive spin 2
representation\foot{If we start from a supersymmetric theory, the
gravitino(s) would also continuously acquire a mass by swallowing
massive spin $1/2$ fields, but this is similar to what happens in
the standard case of deformations which break supersymmetry.}.

Of course, it is not easy to realize this ``Higgs mechanism for
gravity'', since this requires a breaking of the diffeomorphism
symmetry associated to the conserved spin 2 operator $T_{\mu \nu}$.
Up to now, the only known realizations of this phenomenon
\refs{\PorratiGX,\PorratiDB,\DuffWH} were in the context of
``locally localized gravity'' \KarchCT, where an $AdS_{d}$-brane is
embedded into $AdS_{d+1}$ and there is (in some approximation) a
massless bound state of the graviton living on the brane. In the
dual description, there is a $(d-1)$-dimensional defect inside the
$d$-dimensional field theory, and the energy-momentum tensor of the
defect is mapped to the bound graviton. Generally, this
energy-momentum tensor is not exactly conserved since there is an
exchange of energy between the defect and the full CFT, and this
corresponds to the localized massless graviton acquiring a mass
(related to the anomalous dimension of the defect energy-momentum
tensor) \AharonyQF.

If we want to realize this ``Higgs mechanism'' in the full theory on
$AdS_{d+1}$ without breaking the conformal symmetry, it seems like
we need to find a conformal theory in which the energy-momentum
tensor is not conserved, but this is not possible of course.
Instead, we can start, as described in the previous subsection, from
a theory that has more than one conserved energy-momentum tensor,
namely a product of two conformal field theories, and then one of
the symmetries can be broken by coupling together the two conformal
field theories by an exactly marginal deformation without breaking
the overall conformal invariance. From the point of view of the
conformal field theory this precisely realizes the ``Higgs mechanism
for gravity'' described above. When we couple two CFTs together,
each energy-momentum tensor separately is no longer conserved, but
there is still a conserved energy-momentum tensor which is the sum
of the two original energy-momentum tensors plus a contribution from
the deformation
\deformation,
\eqn\tmntot{T_{\mu \nu}^{tot} = T^{(1)}_{\mu \nu} + T^{(2)}_{\mu \nu} -
h \eta_{\mu \nu} {\cal O}_1 {\cal O}_2}
(assuming that the scalar operators contain no derivatives). On the other
hand, the difference between the two energy-momentum tensors acquires an
anomalous dimension since it is no longer conserved : schematically we
have (this will be made more precise below)
\eqn\tmndiff{\del^{\mu} (T^{(1)}_{\mu \nu} - T^{(2)}_{\mu \nu}) \sim
h \left[ (\del_{\nu} {\cal O}_1) {\cal O}_2 - {\cal O}_1 (\del_{\nu}
{\cal O}_2) \right].}
The operator on the right-hand side of \tmndiff\ is the dimension
$d+1$ vector operator which is swallowed by the dimension $d$ spin 2
operator when it acquires an anomalous dimension; note that in this
case it is a multiple-trace operator.

The naive mapping of the CFT results above to anti-de Sitter space
would imply that the product CFT maps to a bulk theory with two
massless gravitons, and then when we deform one of the gravitons
acquires a mass by swallowing a massive vector field, which in this
case is a bound state of two scalar fields (corresponding to the
multi-trace operator on the right-hand side of \tmndiff). This is
indeed what we would find in the ``folded'' interpretation of the
deformed product CFT. In the ``unfolded'' interpretation the
analysis of the fields is somewhat more complicated, since each of
the gravitons (the massless one and the massive one) is a linear
combination of the gravitons on the two AdS spaces. However, since
there is still a single conformal symmetry, the analysis in terms of
representations of the conformal algebra is the same.

When we have a large $N$ limit corresponding to a semi-classical
theory of gravity in the bulk, the graviton mass described above
arises through loop diagrams which are suppressed (at least) by
$1/N^2$ (or by Newton's constant $G_N$). From the bulk point of view
this occurs because the deformation \deformation\ does not directly
affect the graviton, but just modifies the boundary conditions of
the scalar fields which are dual to the operators ${\cal O}_i$. The
leading correction to the graviton propagator thus comes from a
one-loop diagram with the scalar fields running in the loop. On the
field theory side one can easily find the same result by noting that
the first correction to two-point functions of the stress-energy
tensors $T^{(1)}$ and $T^{(2)}$ arises at second order in $h$, and
is suppressed by $1/N^2$ compared to the original two-point
functions.

In the rest of this section, which is rather technical and can be
skipped by readers who are not interested in the details, we will
compute the graviton mass in the product of CFTs deformed by
\deformation\ at leading order in $h$, in the case of a marginal
deformation ($\Delta_1+\Delta_2=d$). We will first compute this on
the field theory side and then on the gravity side, and we will find
precise agreement between the two computations.

\subsec{Field theory computation of graviton mass}

One method to compute the correction to the graviton mass (at
leading order in the deformation \deformation), which is simply
related to the anomalous dimension of the non-conserved
stress-energy tensor, is by using equation \tmndiff\ which relates
its derivative to a specific operator in the undeformed CFT (at
leading order in the deformation). First, we should be more precise
about which operator in the CFT is the non-conserved stress-energy
tensor.

Naively the non-conserved operator which obtains a mass should be
$T^{(1)}-T^{(2)}$, but in fact this operator is not orthogonal (in the
sense of having a vanishing two-point function) to
\tmntot\ and to the operator ${\cal O}_1 {\cal O}_2$.
After the deformation
\deformation, assuming that the deformation operators ${\cal O}_1$
and ${\cal O}_2$ contain no derivatives, it is easy to check (by
applying separate translations of the fields of the two CFTs) that
(at leading order in $h$) \eqn\deftders{\del^{\mu} T^{(1)}_{\mu \nu}
= h (\del_{\nu} {\cal O}_1) {\cal O}_2;\qquad \qquad \del^{\mu}
T^{(2)}_{\mu \nu} = h {\cal O}_1 (\del_{\nu} {\cal O}_2),} such that
the full deformed stress-energy tensor \tmntot\ is conserved. We
will work in the framework of radial quantization, where each local
operator is mapped to a state. In this framework the two-point
functions of operators map to numbers giving the overlaps of states,
proportional to the two-point function after taking out the position
dependence. For scalar operators of dimension $\Delta_i$, with
$\vev{{\cal O}_i(0) {\cal O}_i(x)} = N_i / |x|^{2\Delta_i}$, we
simply have $\vev{{\cal O}_i | {\cal O}_i} = N_i$ for some arbitrary
normalization factor $N_i$. Similarly, for traceless stress-energy
tensors in a conformal field theory we have
\eqn\ttvev{\vev{T_{\mu \nu} | T_{\rho \sigma}} = c
(\eta_{\mu \rho} \eta_{\nu \sigma} + \eta_{\mu \sigma} \eta_{\nu
\rho} - {2\over d} \eta_{\mu \nu} \eta_{\rho \sigma}),}
where $c$ is (one definition of) the central charge of the conformal
theory. We will denote the central charges of the two CFTs (using
this definition) by $c_1$ and $c_2$.

The non-conserved operator must be some linear combination
\eqn\tlincomb{ {\tilde T}_{\mu \nu} = \alpha T^{(1)}_{\mu \nu} +
\beta T^{(2)}_{\mu \nu} + \gamma \eta_{\mu \nu} h {\cal O}_1 {\cal
O}_2.}
Requiring orthogonality to \tmntot\ implies that at leading
order in $h$, $\alpha c_1 + \beta c_2 = 0$. In order to check
orthogonality with ${\cal O}_1 {\cal O}_2$ we need to compute the
leading correction to $\vev{T^{(1)} | {\cal O}_1 {\cal O}_2}$.
Recall that the general prescription for computing corrections to
correlation functions due to the deformation \deformation\ in
conformal perturbation theory is
\eqn\cpt{\eqalign{\langle correlator \rangle_{full} &=
\langle correlator \rangle_{undeformed}  +
h \int d^du \langle correlator\cdot {\cal O}_1(u) {\cal O}_2(u)
\rangle_{undeformed}  \cr
& + {h^2\over 2}\int d^du \int d^dv \langle correlator\cdot {\cal O}_1(u)
{\cal O}_2(u) {\cal O}_1(v) {\cal O}_2(v) \rangle_{undeformed} +\cdots .}}
Thus, we find at leading order \eqn\leadcorr{\vev{T^{(1)}_{\mu
\nu}(x_1) {\cal O}_1(x_2) {\cal O}_2(x_3)} = h \int d^dx
\vev{T^{(1)}_{\mu \nu}(x_1) {\cal O}_1(x_2) {\cal O}_1(x)}
\vev{{\cal O}_2(x_3) {\cal O}_2(x)}.} Using the formulas for
$\vev{T(x_1) {\cal O}(x_2) {\cal O}(x_3)}$ from \OsbornTT\ (which we
quote below) we find that this gives an overlap $\vev{T^{(i)}_{\mu
\nu} | {\cal O}_1 {\cal O}_2} = h {\Delta_i \over d} \eta_{\mu \nu}
N_1 N_2$, which is consistent with the orthogonality of \tmntot\
with ${\cal O}_1 {\cal O}_2$. For the orthogonality of \tlincomb\
with this operator we now require $\alpha \Delta_1 + \beta \Delta_2
+ \gamma d = 0$, so that an appropriate choice of the non-conserved
tensor $\tilde T$ (with an arbitrary normalization) is given by
\eqn\tildetdef{{\tilde T}_{\mu \nu} = c_2 T^{(1)}_{\mu \nu} - c_1
T^{(2)}_{\mu \nu} + (c_1 {\Delta_2\over d} - c_2 {\Delta_1\over d})
\eta_{\mu \nu} h {\cal O}_1 {\cal O}_2.} At leading order in the
deformation, this operator satisfies
\eqn\devtildet{\del^{\mu} {\tilde T}_{\mu \nu} = h (c_1 + c_2)
\left({\Delta_2\over d} (\del_{\nu} {\cal O}_1) {\cal O}_2 -
{\Delta_1\over d} {\cal O}_1 (\del_{\nu} {\cal O}_2) \right).}
As a consistency check, note that if one of the operators (say
${\cal O}_1$) is the identity operator with $\Delta_1=0$, so that
the two theories are not really coupled together by the deformation,
we find that ${\tilde T}$ is still conserved as expected.

Next, we need to compute the relation between the norm of
\devtildet\ and the anomalous dimension of ${\tilde T}$. Recall that
using the conformal algebra, a standard manipulation gives for a
scalar operator $\cal O$ of dimension $\Delta$ (with no summation
over $\mu$) \eqn\dovev{|\del_{\mu} {\cal O}|^2 = |P_{\mu} {\cal
O}|^2 = \vev{{\cal O} | K^{\mu} P_{\mu} | {\cal O}} = \vev{{\cal O}
| 2 i \delta^{\mu}_{\mu} D | {\cal O}} = 2 \Delta \vev{{\cal O} |
{\cal O}},} leading to the standard unitarity bound $\Delta \geq 0$.
The same manipulation for the derivative of a traceless spin 2
operator gives (with no sum over $\nu$)
\eqn\dspintwo{|\del^{\mu} {\tilde T}_{\mu \nu}|^2 =
2 c (\Delta_{\tilde T} - d) {(d+2)(d-1) \over d},}
where $c$ is the constant appearing in \ttvev\ for the 2-point
function of $\tilde T$ with itself, which for ${\tilde T}$ defined
above is equal to $c = c_1 c_2 (c_1 + c_2)$. This equation is
consistent with the known unitarity condition saying that
$\Delta_{\tilde T} \geq d$, with equality if and only if $\tilde T$
is conserved.

Now, we can compare equation \dspintwo\ to the norm of the operator
on the right-hand side of \devtildet. We find
\eqn\forgravmass{2 c (\Delta_{\tilde T} - d) {(d + 2)(d-1)\over d} =
2 h^2 (c_1 + c_2)^2 N_1 N_2 {1\over d^2} (\Delta_2^2 \Delta_1 +
\Delta_1^2 \Delta_2),}
leading to a mass squared of the graviton given by (at leading order
in $h$, and in units of the AdS radius)
\eqn\gravmass{M_{grav}^2 = d (\Delta_{\tilde T} - d) = h^2 N_1 N_2
\left({1\over c_1} + {1\over c_2} \right) {{\Delta_1 \Delta_2
 d} \over (d + 2)(d-1) }.}
Note that despite the appearance of
$N_1$ and $N_2$ this is independent of how we normalize the
operators, since $h$ also changes when we change the normalization
of the two scalar operators. Note also that this scales as the
inverse central charges of the CFTs, namely as $1/N^2$ in the large
$N$ limit of an $SU(N)$ gauge theory, as expected since it is
related to a one-loop diagram in the bulk.

An alternative way to compute the correction to the graviton mass is
by a direct computation of the logarithmic corrections to
$\vev{{\tilde T}(x) {\tilde T}(y)}$, using the techniques of
conformal perturbation theory. We will not do this computation in
complete generality, but we will show that for the case of equal
central charges of the two CFTs it gives results which are
consistent with the previous computation.

The two-point function of $\tilde T$ contains various terms. One term
involves $\vev{T^{(1)} T^{(2)}}$;
the leading order correction to this two-point function is given by
\eqn\ttcpt{\eqalign{& \langle T^{(1)}_{\mu\nu}(x) T^{(2)}_{\sigma\tau}(y)
\rangle_{full} = {h^2\over 2}  \int d^du d^dv \langle T^{(1)}_{\mu\nu}(x)
T^{(2)}_{\sigma\tau}(y) {\cal O}_1(u) {\cal O}_2(u) {\cal O}_1(v)
{\cal O}_2(v)  \rangle_{undeformed} +\cdots } }
Since the two CFTs only interact via the deformation \deformation\
this correlator factorizes into
\eqn\ttfactored{\int d^du\int d^dv \langle T^{(1)}_{\mu\nu}(x)
{\cal O}_1(u){\cal O}_1(v) \rangle^{(1)} \langle T^{(2)}_{\sigma\tau}(y)
{\cal O}_2(u){\cal O}_2(v) \rangle^{(2)},}
where these correlation functions are computed in the respective
undeformed theories.  The general form of these correlation functions
is determined by conformal invariance and is given in \OsbornTT:
\eqn\ttform{\eqalign{\langle T^{(i)}_{\mu\nu}(x) {\cal O}_i(u){\cal O}_i(v)
\rangle^{(i)} &= {A_i \over (x-u)^d (u-v)^{2\Delta_i -d +2} (x-v)^d} \times \cr
& \Biggl( (u-x)_\mu (u-x)_\nu {(x-v)^2\over(x-u)^2} -(u-x)_\mu
(v-x)_\nu \cr & - (v-x)_\mu (u-x)_\nu + (x-v)_\mu (x-v)_\nu
{(x-u)^2\over(x-v)^2} - {1\over d} \delta_{\mu\nu} (u-v)^2\Biggr) } }
where $\Delta_i$ is the conformal dimension of ${\cal O}_i$, $d$ is
the spacetime dimension of the CFT, and $A_i$ is a constant given by
$A_i = - d \Delta_i N_i \Gamma(d/2) / 2 (d-1) \pi^{d/2}$.  Inserting
\ttform\ into the expression \ttcpt\ for the stress-energy tensor
two-point function of interest, and introducing the separation
variable $D^\mu \equiv x^\mu - y^\mu$, we can rewrite the leading
order correction as a sum of 4 independent integrals:
\eqn\ttintegrals{\eqalign{ \langle T^{(1)}_{\mu\nu}(x) &
T^{(2)}_{\sigma\tau}(y)
\rangle_{h^2} = {h^2 A_1 A_2\over 2}
\int {d^du \over u^d} \int {d^d v \over v^d}
{1\over (D-u)^d} {1\over (D-v)^d} {1\over (u-v)^4} \times \cr &
\Biggl\{ \Biggl[ (D-u)_\mu (D-u)_\nu u_\sigma u_\tau
{(D-v)^2\over(D-u)^2} {v^2\over u^2} + (D-u)_\mu (D-u)_\nu v_\sigma
v_\tau {(D-v)^2\over(D-u)^2} {u^2\over v^2} \cr & + \Bigl( (D-u)_\mu
(D-v)_\nu u_\sigma v_\tau + (\sigma \leftrightarrow \tau)
\Bigr) \cr
& - \Bigl( (D-u)_\mu (D-u)_\nu u_\sigma v_\tau {(D-v)^2\over (D-u)^2}
+ (\sigma \leftrightarrow \tau) + \cr & \ \ \ \ \
((\sigma,\tau) \leftrightarrow (\mu,\nu)) +
((\sigma,\tau) \leftrightarrow (\nu, \mu)) \Bigr)  \Biggr]
+ (\mu \leftrightarrow \nu)
- {\rm traces} \Biggr\}.} }
In general, it is quite complicated to regularize and evaluate these
integrals.  Using dimensional regularization, for example, requires
the introduction of four Feynman parameter integrals. However, it is
easy to see that at least some of the terms in \ttintegrals\ diverge
logarithmically, leading to a non-zero anomalous dimension.

The computation of $\vev{{\tilde T}{\tilde T}}$ contains also
various other terms; in particular it contains terms of the form
$\vev{T^{(1)} T^{(1)}}$, and these are difficult to compute at order
$h^2$ since it requires knowing the precise form of $\vev{T^{(1)}
T^{(1)} {\cal O}_1 {\cal O}_1}$ which is not determined purely by
conformal invariance. So, in general we do not know how to compute
the anomalous dimension directly by this method. However, there is a
trick we can use in the special case of $c_1 = c_2$. In this case
${\tilde T} = c_1 (T^{(1)} - T^{(2)} + {{\Delta_2 - \Delta_1}\over
d} \eta h {\cal O}_1 {\cal O}_2)$, and when we compute the
difference $\vev{{\tilde T}(x) {\tilde T}(y) - c_1^2 T^{tot}(x)
T^{tot}(y)}$ the terms proportional to $\vev{T^{(1)}(x) T^{(1)}(y)}$
and to $\vev{T^{(2)}(x) T^{(2)}(y)}$ drop out, and the traceless
part of the resulting expression is simply given by $-4 c_1^2
\vev{T^{(1)}(x) T^{(2)}(y)}$. Since $T^{tot}$ has no anomalous
dimension, the logarithmic terms in this expression should be the
same as the ones appearing in $\vev{{\tilde T} {\tilde T}}$. Even
without computing these logarithmic terms exactly, we can see how
they depend on the operators ${\cal O}_1$ and ${\cal O}_2$, since
the expression \ttintegrals\ depends on these operators only through
the combination $A_1 A_2 \propto \Delta_1 \Delta_2 N_1 N_2$. Thus,
the anomalous dimension arising from the computation above is given
by some function of $d$ times $\Delta_1 \Delta_2 N_1 N_2$, in full
agreement with the result \gravmass\ which we found above.

\subsec{Graviton mass from a bulk gravity calculation}

Since we know how to describe the deformation \deformation\ in terms
of deformed boundary conditions for fields in the bulk, as described
in \S2.1, we can calculate the mass of the graviton in the bulk
directly, using the methods of \refs{\PorratiDB,\DuffWH}, by a
scalar loop correction to the graviton propagator (with the scalars
obeying the deformed boundary conditions). Schematically, one
computes the graviton self-energy in the bulk, and then to find the
mass one extracts the coefficient of the term arising from a massive
spin-1 state by matching the long-distance behavior and the tensor
structure. In \refs{\PorratiDB,\DuffWH} this technique was used to
analyze the mass of the massive graviton that arises in ``locally
localized gravity'' as a bound state on an AdS-sliced brane. Here,
we will use it in the bulk.

The only diagrams that can contribute to the graviton mass at
leading order are those involving the scalars $\phi_1$ and $\phi_2$,
dual to the operators $\co_1$ and $\co_2$.
In the undeformed bulk theory, if we denote the boundary conditions
for the scalars (in the coordinate system $ds^2_{AdS} = z^{-2} (dz^2
+ dx^{\mu} dx_{\mu})$) by $\phi_i(z\to 0) \sim \alpha_i(x)
z^{d-\Delta_i} + \beta_i(x) z^{\Delta_i}$ where $\Delta_i$ is the
conformal dimension of the dual operator $\co_i$ (we assume
$\Delta_i \neq d/2$), the boundary conditions were given by
$\alpha_i(x)=0$.
After the deformation we are considering, the boundary conditions
change to $\alpha_1(x) = h (2\Delta_2 - d)\, \beta_2(x)$ and
$\alpha_2(x) = h (2\Delta_1 - d)\, \beta_1(x)$ : the
``non-normalizable mode'' \foot{Since we are interested in the case
of $\Delta < (d+1)/2$, the mode proportional to $\alpha$ is actually
normalizable, but it is still non-fluctuating in the undeformed
theory because of the boundary condition.} of the scalar in one of
the AdS spaces is related via the deformation to the normalizable
mode of the scalar in the other AdS space. In the scalar loop
diagram contributing to the graviton mass, these altered boundary
conditions create a situation in which the scalar in the loop can
propagate across the boundary from one AdS space to the other and
back.

\fig{The scalar loop graph contributing to the graviton mass in the
two-AdS-space picture. The dashed lines are graviton propagators,
and the solid lines are scalar propagators. The graviton mixes with
a two-scalar state and becomes massive.}{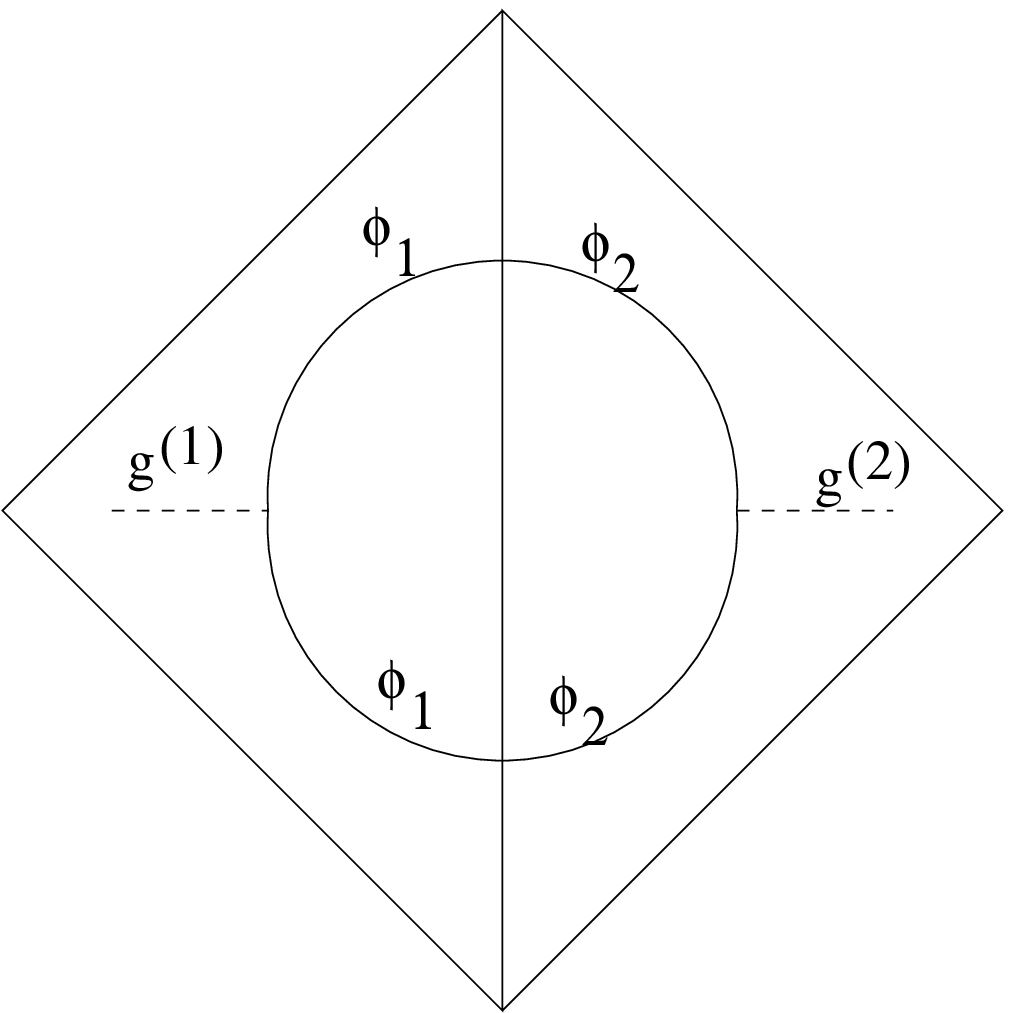}{3truein}
\figlabel{\loopgraph}

In principle we could compute the correction to the graviton mass
directly in this two-AdS-space picture, but in practice it will be
simpler to do the computation thinking about the two gravitons and
the two scalars as living on the same AdS space (of course, this
does not change the physics, but just simplifies the language of the
computation). We will limit ourselves to the case of $c_1=c_2$.
Working in this ``folded'' picture,
we may directly apply the results of \AharonySH\ to find the
corrections to the scalar propagator arising from the deformation
\deformation. The corrected propagator is a $2\times 2$ matrix
since the boundary conditions mix the two scalar fields, and it is
given by \AharonySH\ :
\eqn\scalarprop{G^{ij}_\phi = {1\over 1 + \tilde{h}^2} \pmatrix{
    G^1_\phi + \tilde{h}^2 G^2_\phi & \tilde{h}\, G^1_\phi - \tilde{h}\, G^2_\phi \cr
    \tilde{h}\, G^1_\phi - \tilde{h}\, G^2_\phi & G^2_\phi + \tilde{h}^2
    G^1_\phi}}
where $\tilde{h} \equiv (2 \Delta_1 -d)h$ and $G^i_\phi$ is the
propagator of the $i^{{\rm th}}$ scalar field in the undeformed
theory, namely the propagator corresponding to a field dual to an
operator of dimension $\Delta_i$. The contribution of these scalar
propagators to the graviton propagator (which is now also a $2\times
2$ matrix), through a one-loop diagram in which $g_{\mu \nu}^{(1)}$
couples only to $\phi_1$ and not to $\phi_2$, and similarly for
$g_{\mu \nu}^{(2)}$, are proportional to :
%
 \eqn\gravprop{
    G^{ij}_{\rm grav} \sim {1 \over (1+{\tilde h}^2)^2} \pmatrix{
    (G^1_\phi + \tilde{h}^2 G^2_\phi)^2 & \tilde{h}^2 (G^1_\phi - G^2_\phi)^2 \cr
    \tilde{h}^2 (G^1_\phi - G^2_\phi)^2 & (G^2_\phi + {\tilde h}^2 G^1_\phi)^2 }
.}
As expected, this results in no correction for one linear
combination of the graviton propagators (the sum) and a correction
starting at order $\tilde{h}^2$ to the other (the difference). In
the formalism of \DuffWH, the two point function of stress tensors
must now be promoted to a $2\times 2$ matrix and evaluated carefully
with the corrected scalar propagators.

To extract a graviton mass to match against the field theory result
of the previous subsection we need to compute the correction to the
mass of the ``off-diagonal" graviton.
For the case of $c_1=c_2$ we need to look at the one-loop correction
to the propagator of the graviton dual to $\tilde{T} = (T_1 - T_2) /
\sqrt{2}$ (at leading order in $h$), where for convenience we
normalized $\tilde{T}$ to have the same 2-point function as $T_1$
and $T_2$. This normalization differs from how we defined
$\tilde{T}$ in the field theory, but it is trivial to see that the
normalization of $\tilde{T}$ doesn't affect the graviton mass
(which, in the field theory language, is the scaling dimension of
the operator). The graviton dual to $\tilde{T}$ couples to $\phi_1$
with a positive sign (times $1 / \sqrt{2}$) and to $\phi_2$ with the
opposite sign. Plugging the scalar propagators into the 1-loop
diagram correcting the mass of this specific graviton, one finds
that the diagram is the same as in the unperturbed theory, minus ${
2 {\tilde h}^2 \over (1 + {\tilde h}^2)^2} (G^1_\phi - G^2_\phi)^2$.
Thus, the answer we are looking for is the correction to the
graviton mass coming from a scalar running in the loop with
propagator $G^1_\phi-G^2_\phi$, multiplied (at leading order in $h$)
by $[- 2 h^2 (\Delta_1 - \Delta_2)^2]$. While our field theory
analysis of the previous subsection was valid
only to leading order in $h$, the gravity calculation is done at
leading order in $1/N$ (we only calculate the 1-loop correction),
but it is exact to all orders in $h$ as long as we keep the full
${\tilde{ h}^2 \over (1 + {\tilde h}^2)^2}$ prefactor (equation
\scalarprop\ is exact at leading order in $1/N$). For comparison we
restrict ourselves to the leading term in $h$, but gravity gives us
a prediction for the field theory answer at large $N$ for any $h$,
while the field theory predicts the leading $h$ behavior of the
gravity answer to all orders in $1/N$.

%
In fact, we will now do a more general computation; we will consider
the graviton mass induced by a one-loop diagram of a scalar whose
propagator is $a_{\Delta_1}$ times the propagator of a scalar dual
to an operator with dimension $\Delta_1$, plus $a_{\Delta_2}$ times
the propagator with dimension $\Delta_2=d-\Delta_1$. This is a
generalization of the calculation of \DuffWH\ to arbitrary $d$ and
 $\Delta_1$.
As we just argued, the case we are interested in corresponds to
$a_{\Delta_1} = - a_{\Delta_2} =1$ and has an additional overall
factor of $[- 2 h^2 (\Delta_1 - \Delta_2)^2]$ in the graviton mass.

We need to calculate corrections to the graviton self-energy or the
two-point function, $\langle {\hat T}_{\mu\nu}(x) {\hat
T}_{\rho'\sigma'}(y) \rangle$, of stress-energy tensors in the bulk.
We follow the conventions of \refs{\danbitensors, \AllenWD,
\allenturyn} where unprimed indices indicate tensor indices
evaluated at $x$ while primed indices are evaluated at $y$. Since we
are deforming the boundary conditions of scalar fields, it will
suffice to consider only the scalar field contribution to the bulk
stress tensor $\hat T$. Since AdS is a maximally symmetric
space-time, the stress tensor two-point function may be decomposed
in the following basis of 5 linearly independent bi-tensors
\refs{\DuffWH, \danbitensors, \allenturyn}:
\eqn\dansos{\eqalign{
    \CO_1 & \equiv g_{\mu\nu} g_{\rho'\sigma'}, \cr
    \CO_2 & \equiv \hat{n}_\mu \hat{n}_\nu \hat{n}_{\rho'} \hat{n}_{\sigma'}, \cr
    \tilde{\CO}_3 & \equiv g_{\mu\rho'} g_{\nu\sigma'} + g_{\mu\sigma'} g_{\nu\rho'}, \cr
    \CO_4 & \equiv g_{\mu\nu} \hat{n}_{\rho'} \hat{n}_{\sigma'} + g_{\rho'\sigma'} \hat{n}_\mu
        \hat{n}_\nu, \cr
    \tilde{\CO}_5 & \equiv g_{\mu\rho'} \hat{n}_\nu \hat{n}_{\sigma'} + g_{\mu\sigma'} \hat{n}_\nu
    \hat{n}_{\rho'} + g_{\nu\rho'} \hat{n}_\mu \hat{n}_{\sigma'} + g_{\nu\sigma'} \hat{n}_\mu
    \hat{n}_{\rho'},}}
where ${\tilde \mu}$ is the geodesic distance between $x$ and $y$,
 $\hat{n}_{\alpha(')} \equiv \nabla_{\alpha(')} {\tilde \mu}$
is the unit tangent vector to the
geodesic from $x$ to $y$, evaluated at one end point or the other
according to whether the index is primed, and $g_{\alpha\beta'}$ is
the ``parallel propagator'' which transports unit vectors between
the two points. Rules for manipulating these objects are summarized
in Table 1 of \danbitensors.
To extract the graviton mass following \DuffWH, one must obtain the
constant piece of the coefficient of the transverse-traceless part
of the graviton propagator.  Since this must arise as the effect of
a (bound state) massive spin-1 particle, it suffices to focus on
that tensor structure in the long range behavior of the graviton
self-energy.  First, one computes the full graviton self-energy,
$\Sigma_{\mu \nu \rho' \sigma'} = 8 \pi G_N \langle {\hat T}_{\mu
\nu}(x) {\hat T}_{\rho' \sigma'}(y) \rangle$, by evaluating the
stress tensor two-point function of free scalars and expanding in
the bitensor basis in the limit of large separation of points (large
$\tilde \mu$ or large negative $Z\equiv -\cosh(\tilde \mu)$). This
allows one to neglect contact terms that may arise through use of
the equations of motion to simplify expressions. Second, one
separately computes the transverse-traceless piece which arises due
to the exchange of a massive spin-1 particle (with the appropriate
mass for being swallowed by the graviton, as discussed in \S2.2),
$\Pi^{{\rm spin-1}}_{\mu \nu \rho' \sigma'}$. This may be found by
taking covariant derivatives of the massive spin-1 propagator,
$D_{\mu \rho'}$:
\eqn\vecmass{\Pi^{{\rm spin-1}}_{\mu \nu \rho' \sigma'} = -2
\nabla_{\mu } \nabla_{\rho'} D_{\nu \sigma'}(\mu).}
Matching the coefficient and tensor structure of the leading large
$|Z|$ behavior allows one to identify the mass of the graviton in
this framework, as was done in \DuffWH\ for the special case of a
conformally coupled scalar in AdS$_4$.

We need to perform this calculation for a free scalar field $X$ in
$d+1$ dimensions with Lagrangian
\eqn\scalaraction{  {\cal L} = \sqrt{-g} \left ( -{1 \over 2}
(\partial X)^2 - {m^2 \over 2} X^2 - \alpha {d-1\over 8 d} R X^2
\right ), }
where we allowed contributions to the mass coming either from
explicit masses or from couplings to the background curvature; the
case of $m=0$, $\alpha=1$ corresponds to a conformally coupled
scalar. We normalize the radius of curvature of AdS space to one.
The two possible dimensions of the dual operator are given by
$\Delta_{1,2} = {d \over 2} \pm {1 \over 2} \sqrt{d^2 + 4 m^2 -
\alpha (d^2-1) }$. The case considered in \DuffWH\ corresponds to
$d=3$, $m=0$, $\alpha=1$, $\Delta_1=2$. From the dual field theory
point of view it is clear that the final answer should not depend on
$\alpha$ and $m$ individually, but only on the combination
$\Delta_1$, and we will find that this is indeed the case. The
stress tensor derived from this Lagrangian (after using the scalar
equation of motion) is
\eqn\stresstensor{ \eqalign{ {\hat T}_{\mu \nu} & = (1 - \alpha {d-1
\over 2 d} ) \partial_{\mu} X
\partial_{\nu} X - \alpha {d-1 \over 2d} X \nabla_{\mu}
\partial_{\nu} X  \cr & - g_{\mu \nu} \left ( ({1 \over 2} - \alpha
{d-1 \over 2 d}) (\partial X)^2
 + ({m^2 \over 2} - \alpha m^2 {d-1 \over 2 d} + {\alpha (
\alpha d + \alpha -d) \over 8}
  {(d-1)^2 \over d}) X^2 \right ).  }
}
Calculating the $\vev{{\hat T}{\hat T}}$ two-point function is
straightforward using Wick contractions and taking up to four
covariant derivatives of the scalar propagator. Due to the sheer
number of terms to keep track of, we implemented the rules of Table
1 of \danbitensors\ for manipulations in intrinsic coordinates in
{\tt Mathematica}, and performed the calculation of the correlator,
as well as the subsequent decomposition into tensor structures, on
the computer. For the scalar propagator we use a superposition of
the dimension $\Delta_1$ and $\Delta_2 =(d- \Delta_1)$ propagators
\FreedmanTZ, written using the hypergeometric function $F$ :
\eqn\scalarprop{\eqalign{ G_X(Z) &= a_{\Delta_1} 2^{-\Delta_1}
{\Gamma(\Delta_1) \over \pi^{{d\over2}} (2 \Delta_1 -d)
\Gamma(\Delta_1 -{d \over 2}) } (-Z)^{-\Delta_1} F({\Delta_1 \over
2}, {\Delta_1 +1 \over 2}, \Delta_1-{d \over 2}+1;{1 \over Z^2}) \cr
&+ a_{\Delta_2} 2^{-\Delta_2} {\Gamma(\Delta_2) \over
\pi^{{d\over2}} (2 \Delta_2 -d) \Gamma(\Delta_2 -{d \over 2}) }
(-Z)^{-\Delta_2} F({\Delta_2 \over 2}, {\Delta_2 +1 \over 2},
\Delta_2-{d \over 2}+1;{1 \over Z^2}). }}
For the decomposition into tensor structures we only need the
leading terms in a power series expansion of $\vev{{\hat T}{\hat
T}}$ in large $|Z|$; in order to simplify these expansions in {\tt
Mathematica} we only kept the cross-terms proportional to
$a_{\Delta_1} a_{\Delta_2}$ in the $\vev{{\hat T}{\hat T}}$
correlator, since we expect the mass to vanish in the case where
either $a_{\Delta_1}$ or $a_{\Delta_2}$ vanishes. It is also easy to
see that unless $\Delta_1$ is a half-integer, only cross-terms lead
to integer powers of $Z$ which can match the massive spin-1
structure \vecmass\ we are looking for. For comparison with \DuffWH\
one needs to substitute the notations $a_{\Delta_1} = {1 \over 2}
(a_- - a_+)$, $a_{\Delta_2}=- {1 \over 2} (a_- + a_+)$.

The extraction of the massive spin-1 piece in principle works along
the same lines as in \DuffWH. One new complication in our case is
that while for the conformally coupled scalar considered in \DuffWH,
the $\vev{{\hat T}{\hat T}}$ correlator and hence the self-energy
were transverse and traceless automatically, for the general massive
scalar one first needs to isolate the transverse traceless part. The
self-energy can easily be made traceless by subtracting out the
$\mu$, $\nu$ and the $\rho'$, $\sigma'$ traces. The remainder then
can be decomposed as
\eqn\decomp{ \Sigma^{traceless}_{\mu \nu \rho' \sigma'} =
\Sigma^{\perp}_{\mu \nu \rho' \sigma'} + (\nabla_{\mu} \nabla_{\nu}-
{1 \over d+1} g_{\mu \nu} \nabla^2) (\nabla_{\rho'}
\nabla_{\sigma'}- {1 \over d+1} g_{\rho' \sigma'} (\nabla')^2 )
B(Z)}
for some function $B(Z)$, where $\Sigma^{\perp}_{\mu \nu \rho'
\sigma'}$ is the desired transverse traceless piece that contains
the graviton mass. The decomposition of a general linear metric
fluctuation could also contain a vector piece $\nabla_{\mu} A_{\nu}
+ \nabla_{\nu} A_{\mu}$. A corresponding structure is not present in
our expression for $\Sigma_{\mu \nu \rho' \sigma'}$; presumably this
is due to the fact that the vector piece is not sourced by the
conserved stress tensors. The decomposition \decomp\ was once more
performed using {\tt Mathematica}. The result obtained for the
graviton mass using the (generalized) formalism of \DuffWH\ is
\eqn\mass{M_{grav}^2 = - G_N
{2^{4-d} \pi^{{3 \over 2} - {d \over 2}} \over (d+2) \Gamma( {d + 3 \over 2})}
{
a_{\Delta_1} a_{\Delta_2} \Delta_1\, \Delta_2 \, \Gamma(\Delta_1)
\Gamma(\Delta_2) \over  \Gamma(\Delta_1 -{d \over 2})
\Gamma(\Delta_2 - { d \over 2})},  }
which for $d=3$, $\Delta_1=2$ was previously obtained in \DuffWH\ .

\subsec{Comparison between field theory and gravity}

We are now in a position to put all the bits and pieces together to
compare our result \mass\ for the mass of the graviton with the
field theory result for the anomalous dimension of the non-conserved
stress tensor. As we argued earlier, the case of two CFTs coupled
together via a double trace deformation corresponds to $a_{\Delta_1}
= - a_{\Delta_2} =1$ and has an additional overall factor of $- 2
h^2 (\Delta_1 - \Delta_2)^2$ compared to the case of a single scalar
with mixed propagator we worked out in the previous subsection. So,
the final gravity prediction for the graviton mass is
\eqn\masstot{M_{grav}^2 =  - h^2 (\Delta_1 - \Delta_2)^2 G_N
{2^{5-d} \pi^{{3 \over 2} - {d \over 2}} \over (d+2) \Gamma(
{d + 3 \over 2})}
{ \Delta_1 \, \Delta_2 \, \Gamma(\Delta_1) \Gamma(\Delta_2)
\over \Gamma(\Delta_1-{d \over 2}) \Gamma(\Delta_2 - { d \over 2})}.
}
Note that this expression is positive (even though the expression
\mass\ is not necessarily positive), consistent with our theory
being unitary.

To compare this with our field theory answer \gravmass\ we need to
plug in the appropriate values of $N_1$, $N_2$ and $c$, that is the
normalization of the two-point functions of the operators dual to
canonically normalized bulk scalar fields (as we assumed in the
computation above), as well as the central charge of the field
theory in terms of the bulk Newton's constant. For a canonically
normalized scalar the two-point function of the dual operator with
dimension $\Delta_i$ is \FreedmanTZ\
\eqn\scalartwo{{N_i \over |x|^{2 \Delta_i}}= {1 \over \pi^{d/2}}
{(2\Delta_i -d) \Gamma(\Delta_i) \over \Gamma(\Delta_i- {d \over
2})} {1 \over |x|^{2 \Delta_i}},}
so that for our case with $\Delta_2=d-\Delta_1$ one gets
\eqn\scalarnorm{N_1 N_2 = - (\Delta_1 - \Delta_2)^2
{\Gamma(\Delta_1) \Gamma(\Delta_2) \over \pi^d \Gamma(\Delta_1-{d
\over 2}) \Gamma(\Delta_2 - { d \over 2})}.}
The value of the
central charge may be read from the results of \LiuBU,
\eqn\liutseytlin{c = {1\over 16 \pi G_N}{d (d+1) \Gamma(d) \over
(d-1) \pi^{{d \over 2}} \Gamma ( {d \over 2}) } = {d \over d-1}
{ \Gamma( {d+3 \over 2}) \over G_N  2^{4-d} \pi^{
{3 \over 2} + {d \over 2}}}}
  (in \LiuBU\ the prefactor
of $1/16\pi G_N$ was set to one, but we are reinstating it here).
Plugging these values into  \gravmass, we find an exact agreement
with \masstot, confirming our description of the deformation
coupling the two conformal field theories.

\newsec{Non-duality for non-gauge theories}

In the previous section we argued that some conformal field theories
are not dual to quantum gravity on AdS space (but rather to a sum of
AdS spaces); this naturally raises the general question of
classifying which conformal field theories are dual to quantum
gravity on AdS space, which are dual to a sum of AdS spaces, and
which (perhaps) have no quantum gravity dual at all. In this section
we will discuss the case of free field theories with no gauge
symmetry and the theories which one can flow to from these, and in
the next section we will discuss this question more generally.

Consider a free field theory including some number of free scalars,
fermions and $U(1)$ gauge fields. Can this theory have a quantum
gravity dual ? We claim that the answer is no. A free theory has
many conserved spin 2 operators, which are the energy-momentum
tensors of each free field in the theory, so if it had a dual it
would have to involve many massless gravitons (from the field theory
point of view there is nothing special about the ``total
energy-momentum tensor''). However, each of these spin 2 operators
is not an independent operator, but rather a product (or a sum of
products) of other (gauge-invariant) operators; for a scalar field
we have $T_{\mu \nu} \sim (\del_{\mu} \phi) (\del_{\nu} \phi)$, and
for a free vector field $T_{\mu \nu} \sim \eta^{\rho \sigma} F_{\mu
\rho} F_{\nu \sigma}$. In the AdS/CFT correspondence, we generally
map operators to fields in the bulk, but this is really only true
for a special class of operators (``single-trace'' operators when
the duality involves a large $N$ gauge theory); the states created
by these operators map to single-particle states of the
corresponding fields in the bulk, while the states created by
products of these operators map to multi-particle states in the
bulk\foot{The distinction between these two types of states is not
sharp when the bulk theory is interacting, since they mix together,
but we assume that it still exists.}. If we use this rule for free
field theories, we would conclude that the dual bulk theories (for
each free field) involve a single field in the bulk, which would be
dual to a basic operator (a free scalar field, a free fermion field,
or $F_{\mu \nu}$ for free gauge fields); all other operators in the
theory are products of (descendants of) this basic operator, so they
would map to multi-particle states of this single field in the bulk.
However, this implies that the energy-momentum tensor of the free
field theory would map to a two-particle bound state of this basic
field, and it does not seem likely that this bound state can really
be identified as a massless graviton in the bulk (at least, we do
not know of any consistent examples in which a massless graviton
arises as a bound state, and there are arguments against it in flat
space \WeinbergKQ).

Thus, we claim that free field theories do not map to bulk quantum
gravity theories. This is supported by a well-known example which
arises (for instance) in the near-horizon limit of $N$ D3-branes in
string theory. The low-energy field theory living on $N$ D3-branes
is a $U(N)$ ${\cal N}=4$ super-Yang-Mills theory. This decomposes
into a direct sum of a free $U(1)$ ${\cal N}=4$ theory (including a
vector field, six scalar fields and four fermions) and an
interacting $SU(N)$ theory (at least in the absence of any external
sources in the fundamental representation). Obviously, type IIB
string theory on $AdS_5\times S^5$ does not decompose into a product
of two decoupled theories; and indeed, the spectrum of propagating
fields in this theory seems to map just to the spectrum of operators
in the $SU(N)$ theory, without the $U(1)$ part\foot{So far this has
only been checked in the supergravity approximation, but it is
believed to be true more generally.}. The supergravity spectrum in
the bulk \KimEZ\ does contain so-called ``singleton'' fields (which
are sometimes called ``doubleton'' fields in this case), which are
in a one-to-one correspondence with the free fields of the $U(1)$
multiplet, but these fields can always be gauged away in the bulk.
So, one can think of these fields as living on the boundary; for
most purposes one can just set them to zero and think of the theory
as a pure $SU(N)$ theory, but for other purposes it is sometimes
more convenient to include them (see, for example, \MaldacenaSS).
Clearly, the $U(1)$ theory itself does not have any bulk
gravitational dual (at least not with a propagating graviton).

Based on this example and on the general arguments above, we suggest
that in general free field theories have no gravitational dual, and
that if one couples a free field theory to a gauge theory with a
gravity dual, one should think of this free field theory as living on
the boundary. Once we accept this for free field theories, it seems
that it must be true also for any deformations of free field theories,
since it is hard to imagine how a bulk theory would suddenly emerge
when we deform. So, we claim that any deformation of a free field
theory, such as the infrared fixed point of the $\phi^4$ field theory
of a single scalar field in $2+1$ dimensions, also has no gravity
dual. Note that we are discussing here deformations by adding
gauge-invariant operators to the Lagrangian; by such deformations we
cannot obtain a theory with non-trivial gauge interactions.
%
It is easy to see that the key feature of the stress tensor failing
to be an independent operator survives such a deformation. Turning
on non-gauge interactions will add terms to the stress tensor, but
the new terms will also be products of gauge-invariant operators,
not independent operators.

Which theories can have a gravity dual ? In all known examples which
have an explicit Lagrangian description, these theories have a gauge
group which acts non-trivially on {\it all} the fields.  Then, the
energy-momentum tensor can no longer be written as a product of
gauge-invariant operators (it is generally a sum of products of
non-gauge-invariant operators), and it makes sense to identify it
with a massless bulk graviton. The gauge group can be continuous (as
in $SU(N)^k$ gauge theories with bifundamental and adjoint fields)
or discrete (as in the $1+1$ dimensional sigma model on the
$T^{4N}/S_N$ orbifold). In some cases the gauge theory could have a
free field limit where it does not contain any interactions (for
instance, in the ${\cal N}=4$ $SU(N)$ super-Yang-Mills theory one
could take the $g_{YM} \to 0$ limit), but even in this limit it is
not identical to a free field theory of the type discussed above,
since the path integral still involves a division by the gauge
symmetry, so (at least on a compact space) the spectrum of a free
$SU(N)$ theory is very different from the spectrum of a free
$U(1)^{N^2-1}$ theory.

Does every gauge theory have a quantum gravity dual ? We do not know
the answer to this question, but it seems likely that the answer is
positive; many theories can be reached by flows from known examples,
and there is no natural separation of gauge theories into two
classes (one which would have a gravity dual and one which would
not). Then,
%
when we couple additional singlet fields to a gauge theory, these
fields would live on the boundary rather than in the bulk (this has
a simple description in the AdS/CFT correspondence, by just making
the coupling constants of the CFT, which have a known description as
boundary conditions in AdS space, dynamical). However, our
discussion of the previous section suggests that the story is more
complicated, and that some gauge theories are actually dual to a
quantum gravity theory on a sum of several spaces; we will discuss
this in the next section.


\newsec{The ``connectivity index''}

The discussion of section 2 suggests that the space of conformal
field theories can be classified according to a ``connectivity
index'' $n$, labeling the number of components in the gravity dual
space-time (which are only connected at their boundary). As
discussed in the introduction, it is not clear that this index is
well-defined (in the sense that theories with different $n$'s could
not be identified in some way), but it certainly seems to be
well-defined in the semi-classical approximation, and it is hard to
see how quantum corrections could change the number of components of
space-time (near the boundary), so we will assume here that it is
well-defined, and see if we can understand this from the field
theory point of view.

The discussion of the previous sections suggests that the
``connectivity index'' $n$ is related to the number of independent
gauge groups, and that it can be defined in the following way :
given a Lagrangian formulation of a field theory in terms of a gauge
group $G$ (continuous or discrete), then $n$ is the maximal number
such that we can write $G = G_1 \times G_2 \times \cdots \times
G_n$, and such that no field is charged under more than one $G_i$
factor.

The definition above raises several immediate questions. One
disturbing issue is that it involves not only having a Lagrangian
formulation for the theory, but also a counting of gauge groups,
even though gauge groups are not really physically well-defined
objects but just redundancies in our description of a theory.
Obviously, if $n$ is really well-defined, there should be a way to
define it which does not refer to a counting of gauge groups or to a
Lagrangian formulation, but just to abstract properties of the
theory. So far we have not been able to find such a more general
definition, but we believe that it exists. Clearly, a theory with
connectivity index $n$ should have at least $n$ spin 2 operators
which cannot be written as (sums of) products of other operators,
and which can be identified with (linear combinations of) the
gravitons on the $n$ components of space-time. However, generally
all but one of these spin 2 operators have dimensions bigger than
$d$, and generic theories have many such operators, so that it is
difficult to identify which of the spin 2 operators correspond to
gravitons and which do not (the question may not even make sense
beyond the semi-classical gravity limit). In examples of the type we
discussed in section 2, we can deform the parameters of the field
theory with index $n$ continuously to a point where it is a product
of $n$ independent theories (with $n$ independent spin 2 operators
of dimension $d$), and then it is clear that the index must be at
least $n$; but it is not clear if it is always possible to do this
for any theory with $n>1$.

Note that the value of $n$ can change when we take a low-energy (IR)
limit of a non-conformal field theory. Clearly, $n$ can decrease in
the IR if we have a product of some gauge groups, and some of them
confine and develop a mass gap. In the gravity dual this would mean
that some of the $n$ spaces do not host any low-energy fields, so
that the IR limit involves only the other spaces. It is also
possible for $n$ to increase as we flow; a simple example of this
which we mentioned above is the point on the moduli space of the
${\cal N}=4$ SYM theory where $SU(N)$ is spontaneously broken to
$SU(N_1)\times SU(N-N_1)\times U(1)$. Despite the product structure,
this theory has $n=1$ (since there are bifundamental fields), but as
we flow to the IR it decouples into a product of three theories, two
which have a gravity dual (which is just a smaller $AdS_5\times
S^5$) and one (the $U(1)$ theory) which does not. Again, this has a
simple picture in the gravity dual, as we discussed above; we have a
flow which in the UV is given by a single $AdS_5\times S^5$ space,
but there are two ``throat'' regions in this space where low-energy
fields live. Each of these regions locally looks like $AdS_5\times
S^5$, and as we go to energies below the scale of the string
stretching between the two ``throats'', we get two decoupled
theories (which in some sense share the same boundary where the
``throats'' connect).

Let us consider two more examples. The theory of $SU(N_c)$ SQCD with
$N_f$ flavors and $N_c \leq N_f < 3N_c/2$ is believed to flow to a
free theory in the IR. For $N_f \leq N_c+1$ this is a free theory of
scalars and fermions, so we have a flow from $n=1$ to $n=0$, while
for $N_f > N_c+1$ the IR theory is a free $SU(N_f-N_c)$ gauge
theory, so we have a flow from $n=1$ to $n=1$. For $3N_c/2 < N_f <
3N_c$ the theory flows to an interacting superconformal field
theory, which is believed to also be the end-point of the flow from
the dual ``magnetic'' $SU(N_f-N_c)$ theory \SeibergPQ. In this case
it seems that on both sides of the duality we have a flow from $n=1$
to $n=1$, since it seems unlikely that the energy-momentum tensor in
the IR SCFT would be a composite operator. Note that the
``magnetic'' theory has many singlet fields, and in the dual
gravitational description we argued that they should be interpreted
as living on the boundary of space-time.

Another example is the $d=3$ ${\cal N}=2$ supersymmetric $U(1)$ gauge
theory with one positively charged and one negatively charged chiral
multiplet ($N_f=1$). This theory is believed \AharonyBX\ to flow to
the same non-trivial IR fixed point as the theory of three chiral
multiplets with a superpotential $W = X Y Z$. So, in this case we have
flows to the same fixed point from theories with $n=0$ and with $n=1$.
It seems likely that the IR theory in this case has $n=0$, since it
includes $X$, $Y$ and $Z$ as primary operators, and the energy-momentum
tensor is a composite in these variables.

According to our conjectures the $O(N)$ vector model in $d=3$ should
not have any quantum gravity dual (it has $n=0$), even in the large
$N$ limit; however, when the $O(N)$ symmetry is gauged (even very
weakly), such that only the singlet sector of this model is
physical, the model (with $n=1$) could have a quantum gravity dual
(as suggested in \KlebanovJA, see \ElitzurKZ\ for a recent discussion).

Unfortunately, it seems that even with all the caveats above the
definition that we gave is too naive. This is because\foot{We thank
D. Kutasov for reminding us of this fact.} there is a
counter-example where we can connect a theory with $n=0$ to a theory
with $n=1$ by a marginal deformation (which should not change $n$
according to our arguments); this is the example of the $c=1$
conformal field theory of a free scalar field on a circle, which is
connected by marginal deformations to the theory of a scalar on
$S^1/Z_2$. According to our general arguments, the first theory
should have $n=0$ and the second (involving gauging a discrete
group) should have $n=1$. Thus, it seems that our definition above
is too naive (at least when low dimensions and discrete gauge groups
are involved), and that it should be made more precise. We hope that
there exists some direct and precise field theory definition of $n$,
but we have not yet been able to find it.

\newsec{Summary and conclusions}

In this paper we constructed the natural generalization of the
AdS/CFT correspondence to product CFTs deformed by multi-trace
operators. The natural gravitational dual to a product of $n$ CFTs
is quantum gravity on $n$ AdS spaces, with their boundaries
identified in the sense that the boundary conditions of fields on
one AdS space are related to the boundary conditions of fields in
the other AdS spaces. These altered boundary conditions will
generically create a situation where one linear combination of the
bulk gravitons remains massless while all other linearly independent
combinations acquire some mass proportional (at leading order in the
deformation, but valid for any conformal field theory, independent
of the large $N$ limit) to the square of the deformation parameter,
$h$, via the AdS Higgs mechanism for gravitons.  For special cases
when the CFTs are identical, the bulk picture can be ``folded'' and
thought of as a single AdS-space with one massless and several
massive gravitons. For more general theories, the unfolded picture
is necessary to accommodate theories whose gravitational duals have
different compact spaces or which are different in the IR (for
instance, one could consider a product of a conformal gauge theory
with a confining gauge theory).

We have made this construction explicit in the semi-classical limit
(which, for gauge theories, is the same as the 't Hooft large $N$
limit), in the case of a product of field theories which admit
semi-classical gravitational duals. Our construction suggests the
existence of a ``connectivity index'' characterizing field theories,
which on the gravity side counts the number of components of
space-time, and on the field theory side roughly counts the number
of independent gauge groups.  We have only explored this in detail
in the semi-classical limit. However, since it is difficult to
imagine that quantum effects could change the integer number of
components of space-time which share a common boundary, we
conjecture that this integer is in fact a true index, and that a
general definition of this index can be found for field theories,
which does not rely on a Lagrangian or gauge theory formulation.  We
conjecture that this index could then be used to classify field
theories according to whether they have a gravitational dual
description in the following way : field theories with $n=0$ {\it
will not} have a dual description as quantum gravity on some
asymptotically anti-de Sitter space, while field theories with $n
\geq 1$ {\it will} have a dual description as quantum gravity on $n$
asymptotically anti-de Sitter spaces with a common boundary, in the
manner discussed above.


We have discussed in detail only the case of coupling together
conformal theories in a way which preserves conformal invariance,
but the generalizations to many other cases are straightforward. For
instance, it is easy to discuss the case of coupling together two
theories at finite energy density or temperature, by replacing the
AdS backgrounds by AdS black holes. When the two theories have the
same temperature the system is in thermal equilibrium, while
otherwise there will be a flow of energy from one theory to the
other across the boundary (which will be very slow in the large $N$
limit). It may also be interesting to put in a finite UV cutoff
(integrating out the high-energy modes), in which case our model
becomes a version of the two-throat Randall-Sundrum \RandallVF\
model (involving two field theories coupled to gravity). 
In the presence of a finite cutoff, one could also couple directly the
stress-energy tensors of the two CFTs, and obtain a graviton mass of
order the curvature scale (rather than a one-loop mass suppressed by
$G_N$ as in our discussion), as discussed in a similar context in 
\GabadadzeJM.

We did not discuss here the string theory construction of the
theories dual to deformed product CFTs, but the procedure to obtain
this is a straightforward generalization of the discussion in
\AharonyPA. The string description of a product of CFTs is by a
worldsheet theory which is a sum of two sigma models (on a product
of AdS space with some compact space), such that each connected
component of the worldsheet maps to one of the two space-times. The
deformation couples the two sigma models together, but this coupling
is non-local on the worldsheet, and given by the translation to the
worldsheet of \deformation\ (with each space-time operator mapping
to an integrated vertex operator on the worldsheet).

\medskip

{\bf Note added:} We are aware that E. Kiritsis has independently
been working on similar ideas \KiritsisHY.

\medskip

\centerline{\bf Acknowledgements}

We would like to thank A. Adams, D. Berenstein, S. Cherkis, O.
DeWolfe, E. Kiritsis, D. Kutasov, J. Liu, J. Maldacena, A. Maloney,
E. Martinec, T. Petkou, N. Seiberg, S. Shenker, and M. Strassler for
useful discussions and correspondences. OA would like to thank
Stanford University, SLAC, the Aspen Center for Physics, and the
Institute for Advanced Study for hospitality during the course of
this work, and the organizers of the Eurostrings 2006 conference for
the chance to present the results of this work there (see
\ofertalk).  The work of OA was supported in part by the Israel-U.S.
Binational Science Foundation, by the Israel Science Foundation
(grant number 1399/04), by the Braun-Roger-Siegl foundation, by the
European network HPRN-CT-2000-00122, by a grant from the G.I.F., the
German-Israeli Foundation for Scientific Research and Development,
by Minerva, and by a grant of DIP (H.52). The work of ABC and AK was
supported in part by DOE contract \# DE-FG02-96-ER40956.

\listrefs

\end